\documentclass[sn-aps,iicol]{sn-jnl}%
\RequirePackage{hyperref}
 
\usepackage{graphicx}%
\usepackage{multirow}%
\usepackage{amsmath,amssymb,amsfonts}%
\usepackage{amsthm}%
\usepackage{mathrsfs}%
\usepackage[title]{appendix}%
\usepackage{xcolor}%
\usepackage{textcomp}%
\usepackage{manyfoot}%
\usepackage{booktabs}%
\usepackage{algorithm}%
\usepackage{algorithmicx}%
\usepackage{algpseudocode}%
\usepackage{listings}%
\usepackage{svg}
\usepackage{subcaption}
\usepackage{siunitx,upgreek}
\usepackage{booktabs}
\usepackage{float}

\newcommand{\blfootnote}[1]{%
    \begingroup
    \renewcommand{\thefootnote}{}%
    \footnote{#1}%
    \addtocounter{footnote}{-1}%
    \endgroup
}

\begin{document}

\title[Article Title]{Asymmetry of Imploding Detonations in Thin Channels}

\author*[1]{\fnm{Sebastian} \sur{Rodriguez Rosero}}
\email{sebastian.rodriguezrosero@mail.mcgill.ca}

\author[2]{\fnm{Jason} \sur{Loiseau}}

\author[1]{\fnm{Andrew J.} \sur{Higgins}}

\affil[1]{\orgdiv{Department of Mechanical Engineering}, \orgname{McGill University}, \orgaddress{\street{817 Sherbrooke St W}, \city{Montreal}, \postcode{H3A 0C3}, \state{Quebec}, \country{Canada}}}

\affil[2]{\orgdiv{Department of Chemistry and Chemical Engineering}, \orgname{Royal Military College of Canada}, \orgaddress{\street{13 General Crerar Crescent}, \city{Kingston}, \postcode{ K7K 7B4}, \state{Ontario}, \country{Canada}}}

\abstract{The factors that influence the symmetry of an imploding detonation are investigated experimentally and theoretically. Detonations in sub-atmospheric acetylene-oxygen were initiated and made to converge in an apparatus that followed that of Lee and Lee (Phys. Fluids \textbf{8}:2148–2152 (1965)). The width of the test section was controlled with a wave-shaping insert, which formed the test section against the viewing window, creating an effectively two-dimensional problem with a channel width comparable to the detonation cell size. The convergence of the detonation was observed via self-luminous open-shutter photography and high-speed videography. The resulting videos were analyzed to quantify the wave speed, degree of asymmetry, and direction and magnitude of the offset in the center of convergence. To determine the experimental parameters that influence the symmetry of the imploding wave, the wave-shaping insert was intentionally canted by 0.3--0.6°, accentuating the asymmetry of the imploding detonation. The experiment was modeled using a Huygens construction wherein the detonation is treated as a collection of wavelets, each assumed to propagate locally at a velocity determined by the channel width. The results of the model reproduced the observed offsets in detonation convergence from the center of the apparatus, confirming that velocity deficits resulting from the narrow channel width control the observed asymmetry.}


\maketitle

\blfootnote{This version of the article has been accepted for publication, after peer review, but is not the Version of Record and does not reflect post-acceptance improvements, or any corrections. The Version of Record is available online at:\url{https://doi.org/10.1007/s00193-024-01196-z}.}

\section{Introduction} 
\label{sec1}

Due to the potential to attain high-energy-density states with minimal energy input, imploding shock waves and detonations remain a fascinating phenomenon. For a perfectly symmetric imploding detonation or shock wave, a singularity in temperature and pressure is theoretically obtained at the center of implosion \cite{Guderley1942}, while in practice asymmetries resulting from instability limit the states that may be obtained \cite{Gardner1982}. Applications of imploding detonations include serving as effective pre-detonators for fuel-air engines \cite{jacksonPDE} and potentially as initiators for fusion energy reactors \cite{Nuclear_Detonations}. Imploding detonations have also been demonstrated to be an effective metal-shaping mechanism \cite{hondaSheetMetalForming1999}.

The potential usefulness of the focusing phenomenon makes it of interest for study in various fields, with most experiments done with imploding shocks (rather than detonations). First, the seminal work of Perry and Kantrowitz in 1951 investigated whether converging shock formation was possible in a cylindrical shock tube with a converging-diverging obstacle \cite{Perry_Kantrowitz}. Later, Saito and Glass revisited imploding shocks with an explosive-driven hemispherical cavity \cite{Saito_Glass}. Through spectroscopic measurements, their study revealed a temperature of more than 10,000~K at the focal point. More recently, others have revisited the converging shock wave problem, particularly on forming complex polygonal shocks through various obstacle configurations \cite{Kjellander}. A comprehensive review of imploding shock research can be found in the recent book by Eliasson and Apazidis \cite{Shock_focusing_book}.

\begin{figure}[htbp]
    \centering
    \includegraphics[width=0.99\columnwidth]{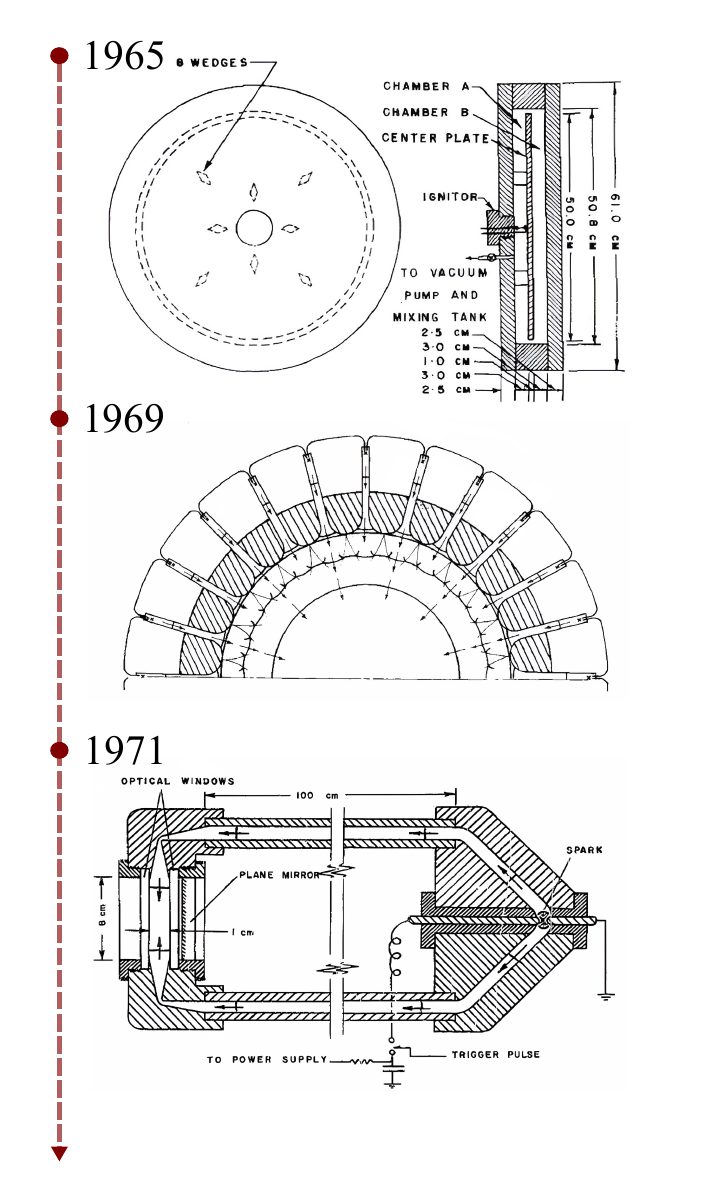}
    \caption{Three generations of imploding detonation experiments were done by Lee and Knystautas, graphics from \cite{leeCylindricalImplodingShock1965, knystautasSparkInitiationConverging1967, knystautasDiagnosticExperimentsConverging1969, Lee_3}.}
    \label{fig: Lee_3_generations}
\end{figure}

In contrast with imploding shock waves, the problem of imploding gaseous detonations has been comparatively less experimentally studied. The development of an experimental apparatus to investigate imploding detonations was performed by Lee and Knystautas in the 1960s. In their study, three generations of the imploding detonation experiment were tested, as seen in Fig.~\ref{fig: Lee_3_generations}. The first, in 1965, consisted of a series of stacked flanges with a center disk held at the center of a hollow cavity. Here, the detonation would initiate on one side of the disk, wrap around, and converge on the other side, where a window was located \cite{leeCylindricalImplodingShock1965}. This study served to experimentally validate the self-amplifying characteristics of imploding detonations, a phenomenon previously described using the Chester-Chisnell-Whitham (CCW) method \cite{Whitham1999}. Their second generation, from 1969, changed how the implosions were formed. This apparatus contained a wide cylindrical cavity with an array of tubes attached to the outer wall. A series of initially planar detonations would be spark-initiated at the end of the tubes; these hemispherical expanding detonations would then enter the cylindrical cavity and combine to form a polygonal converging detonation \cite{knystautasSparkInitiationConverging1967, knystautasDiagnosticExperimentsConverging1969}. This new apparatus improved the optical access to the experimental test section. In 1971, their third and last version of the experiment was constructed. This iteration used a long annular channel to create a planar detonation, which then propagated through a converging-diverging section before reaching the test section, where a double window arrangement allowed for observation \cite{Lee_3}. In this last study, the presence of a curvature preservation mechanism that acts against local wavefront perturbations was first hypothesized.

Ahlborn and Huni performed further investigation of imploding detonations in \cite{ahlborn}, where the velocity profile of an imploding detonation was quantified and was found in agreement with the results reported in Lee and Lee \cite{leeCylindricalImplodingShock1965}. Later, Terao and co-workers revisited the imploding detonation problem in \cite{Terao_1, Terao_2, Terao_3} where they confirmed the existence of self-amplification of the converging front in the case of spherically converging detonations. Recent studies of imploding detonations in condensed-phase explosives have adopted a similar approach to Knystautas et al. in \cite{knystautasSparkInitiationConverging1967} to evaluate different aspects of multipoint initiation \cite{sultanovFormationConvergingDetonation2023,dudinExperimentalInvestigationCylindrical2016a,sosikovSmoothingFrontDetonation2019a}.

As described in \cite{Lee_3}, the degree of symmetry in which a detonation wave converges directly determines the extent to which energy is accumulated at the focal point. Thus, in theory, a perfectly symmetric imploding detonation would lead to an infinitely dense energy state. However, it was also mentioned that although, ``there is no obvious physical characteristic of the apparatus that could be traced to cause the particular asymmetries observed,'' it was found to be challenging to create perfectly symmetric imploding detonations. At the time, it was hypothesized that this asymmetry originated from instabilities inherent in the detonation front, which persisted to the point of focus of the wave. However, the exact mechanism that causes the asymmetry has yet to be explicitly identified. Therefore, this study aims to revisit the problem of imploding cylindrical detonations in gaseous mixtures, particularly to provide new insight into the primary mechanisms that cause asymmetry in imploding cylindrical detonations. 

To avoid complex three-dimensional effects, the dynamics of the wave across the thickness of the cylindrical channel can be suppressed by making the channel thin, i.e., with a thickness comparable to the detonation cells size. Recently, the use of thin channels—conceptually similar to Hele-Shaw cells in classical fluid dynamics—by Radulescu and co-workers has permitted the suppression of cellular dynamics in the direction perpendicular to the visualization window, enabling the dynamics of two-dimensional cellular detonations \cite{radulescu2018, Xiao2020a, Xiao2020b, Xiao2021} and shock-flame interactions \cite{yang2021} to be revealed in exquisite detail. Adopting the methodology of a thin channel, however, can result in the dynamics of the wave being extraordinarily sensitive to the geometric details of the experimental apparatus, particularly in an imploding geometry. Quantification of this sensitivity is the objective of the present study.

In section \ref{sec2} of the present paper, the imploding detonation apparatus is presented in detail. The use of open-shutter photography and high-speed videography of the imploding detonations is presented in sections \ref{sec3} and \ref{sec4}, respectively, and metrics to quantify the symmetry of the imploding wave are proposed. To investigate the mechanism for the observed sensitivity of the imploding detonation, the apparatus is intentionally operated with a slight gradient in channel width across the diameter of the cylindrical test section. The results are analyzed and modeled via a Huygens-like construction of the detonation wave dynamics in section \ref{sec5}, and the comparison of this model to the experimental results is discussed in section \ref{sec6}.

\section{Experimental device} \label{sec2}

\begin{figure*}[!t]
    \centering
    \includegraphics[width=\textwidth]{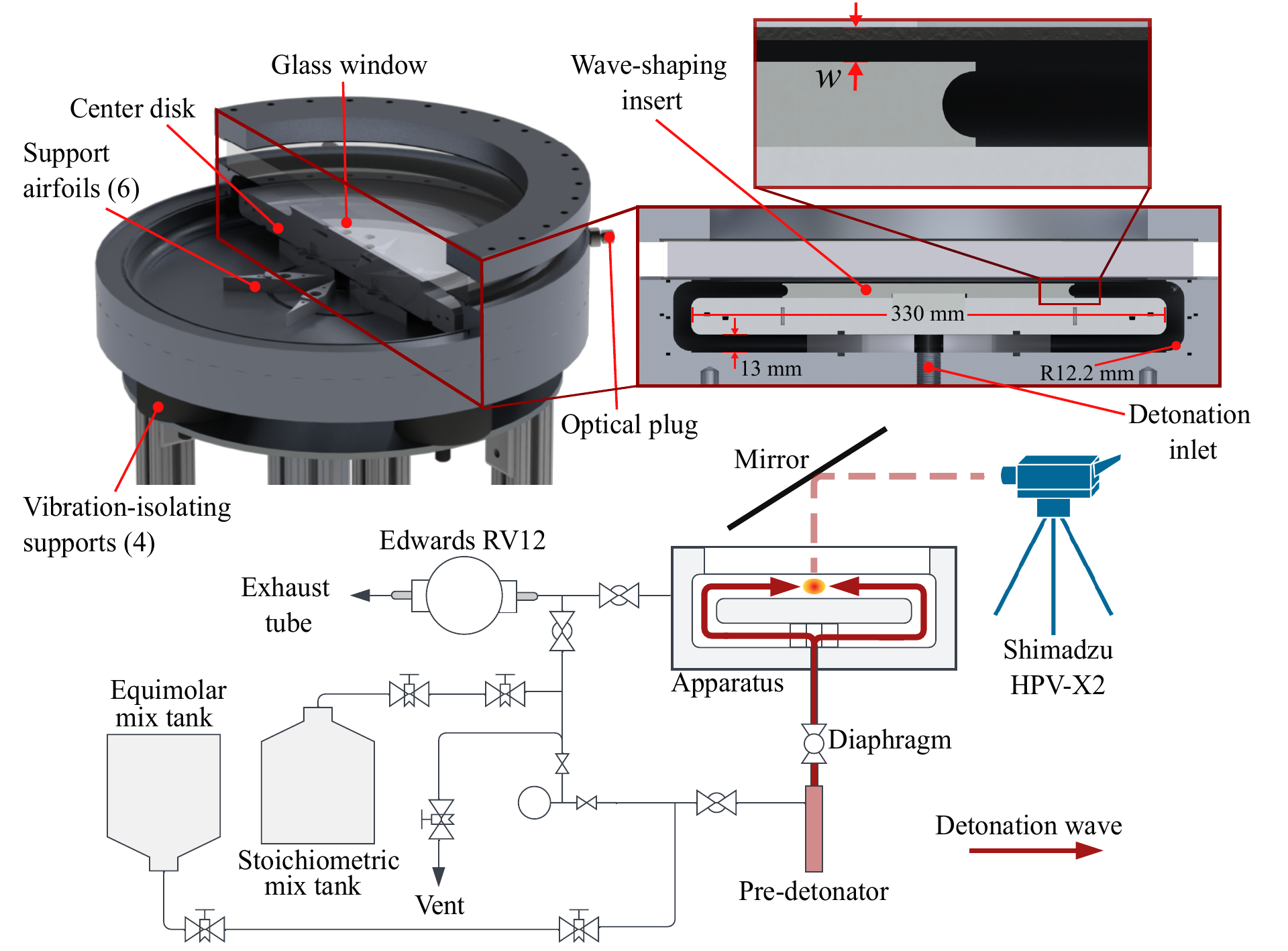}
    \caption{Section and detailed views of the experimental apparatus displaying key features and dimensions. A plumbing and instrumentation diagram is also included in the figure. At the top of the figure, $w$ is the channel width at the test section.}
    \label{fig: device}
\end{figure*}

The implosion device consisted of a series of aluminum stacked flanges assembled to create an internal cylindrical cavity with rounded edges, measuring 50.8~mm in height and 359.7~mm in width. A 25-mm-thick, 330-mm-wide center disk created the constant-width 13~mm channel through which the detonations propagated. Figure~\ref{fig: device} shows a section view of the experimental apparatus, highlighting its main features, the gas handling system used, and the path through which the detonations propagate before convergence. As seen in Fig.~\ref{fig: device}, the detonations originated at the pre-detonator (described below) and propagated through a 12.5-mm-inner-diameter tube until entering the cavity through a 19.05-mm-diameter inlet, wrapping around the center disk, and converging at the test section. Two different setups were used to capture the imploding detonations: First, a primary surface mirror mounted above the apparatus was used with a Shimadzu HPV-X2 camera (CCD sensor resolution of 400$\times$250~px, triggered with a PDA-55 photodiode) to record high-speed videos. Second, a Sony Alpha~6000 DSLR camera was mounted directly above the window to capture open-shutter pictures of the detonations while performing experiments in a darkened laboratory. For some of the open-shutter pictures, a Thorlabs 550-nm short-pass filter (FESH0550) was used to limit the light captured from the reaction products. Two window designs were tested, the first with a small 50-mm-wide window in an aluminum flange, and the latter with a larger 304.8-mm-wide window that spanned the entire test section, as shown in Fig.~\ref{fig: device}. Only the experiments using the large window were used for the analysis reported in this paper. Some of the early tests done with the small window are included in the supplementary material.

Since the symmetry of the imploding detonations was influenced by the concentricity of the flanges and center disk, all parts were machined with 0.001-inch (25 µm) tolerance using interlocking lips. In addition, because the apparatus was sealed using Buna-N rubber o-rings and gaskets (for the window), the 18 grade-8 bolts that held the apparatus together had to be uniformly torqued to a set amount, typically around 3.5~$\mathrm{N\cdot m}$, to ensure a uniform channel width. This technique was adopted as it was noticed through early experiments that slight changes in the assembly of the apparatus would lead to significantly different symmetry of the imploding detonation. Controlling the compression of the different seals greatly improved the reproducibility of the experiment. For each set of tests, the width of the test section was measured using a micrometer and a reference plane placed on top of the top ring, before closing the apparatus, which was kept sealed until all tests were finished. The compressed thickness of the 1.59-mm-thick rubber gasket sealing the window was then added to the measured width. A value of $20\%$ compression was estimated from the measured data and used consistently for all tests. Because the apparatus was not modified between tests, any asymmetry originating from the assembly should be consistent for tests of the same kind.

Two configurations for supporting the center disk were tested: (1) a support cylinder with six diverging channels and (2) six individual diamond-shaped airfoils with extended trailing edges; the second configuration was used for all tests reported in this paper. A brief discussion of the effects observed while using different support designs is included in the Appendix. 

At the point of convergence, a modular aluminum insert sat on top of the center disk and was used to reduce the channel width at the test section to a few millimeters. This \emph{wave-shaping insert} was used to reduce three-dimensional effects by trimming the detonation front, allowing only a thin wave layer into the converging section. Two inserts with different diameters, 204 and 254~mm, were used; both inserts have rounded walls meant to redirect the discarded section of the detonation wavefront away from the test section, limiting its influence on the resulting implosion. In addition, the inserts were sandblasted and painted matte black to improve the visibility of the wavefronts. 

Detonation initiation was done through spark discharge with a custom-made energy delivery system. Since the spark energy was insufficient to consistently achieve direct initiation or prompt DDT at low initial pressure (below 15~kPa), a pre-detonator section was installed. This section consisted of a 12.5-mm-wide and 381-mm-long cavity filled with a sensitive gas mixture at a much higher initial pressure than the inside of the apparatus. A Shchelkin spiral was added to promote the fast occurrence of DDT within this section. The pre-detonator was filled to 100~kPa and separated from the rest of the experiment by a 12.7-$\upmu$m-thick mylar diaphragm which was ruptured by the high-pressure wave after initiation.

The gaseous mixtures used are stoichiometric and equimolar acetylene-oxygen. While the stoichiometric mixture was used for experiments, the more sensitive equimolar mixture was used for the pre-detonator. Acetylene-oxygen was chosen as the test mixture because of its small cell size at the pressure range that the experimental apparatus could tolerate (2.5 to 40~kPa). This mixture was also easier to initiate than other gaseous mixtures, such as ethylene-oxygen or hydrogen-oxygen, which were also tested. Both mixtures were prepared in separate mixing tanks via the method of partial pressures using the gas-handling system shown in Fig.~\ref{fig: device}, and several days were allowed before testing to ensure the proper mixing of gases inside the tanks. The apparatus was evacuated between tests to the lowest measurable pressure for the experiment, $\approx0.1$~kPa, for several minutes to remove residual gas.

\section{Open-shutter photography}\label{sec3}
\begin{figure*}[p]
    \centering
    \includegraphics[width=0.8\textwidth]{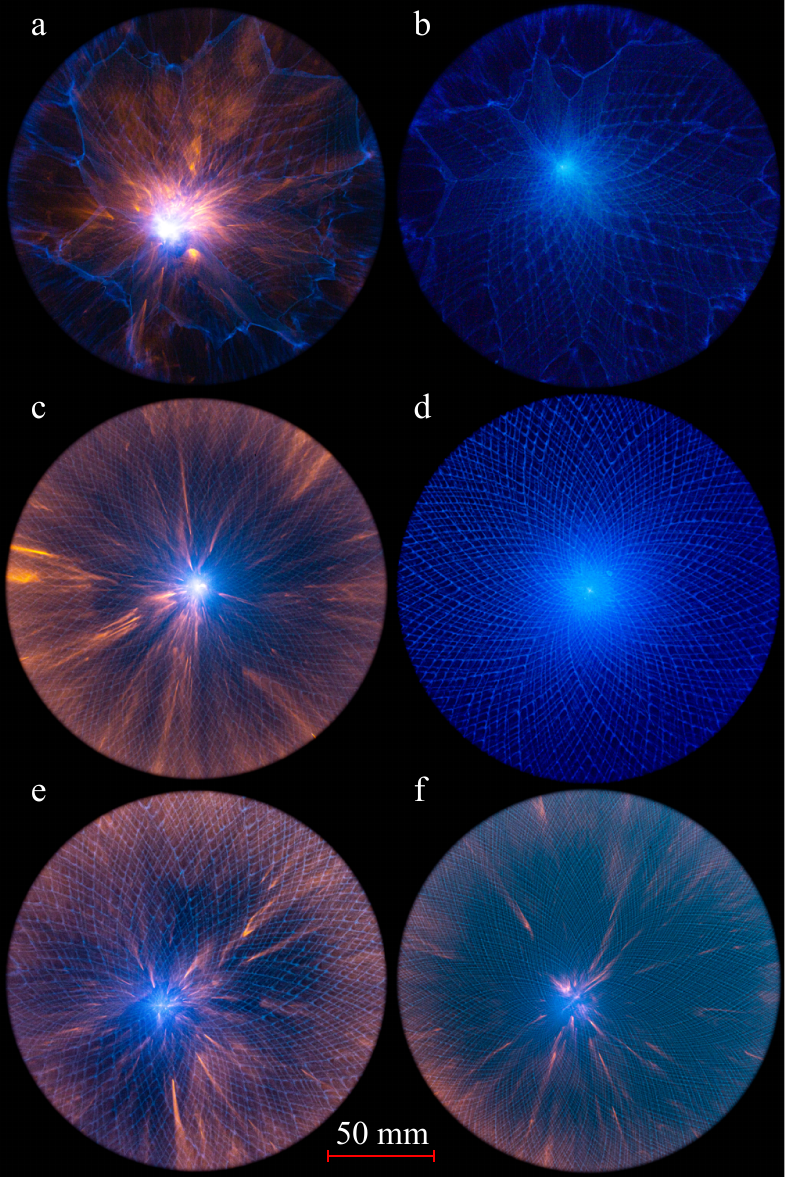}
\caption{Showcase of different open-shutter pictures captured with the experimental apparatus. Pictures \textbf{a} and \textbf{b} captured at 2.5~kPa, \textbf{e} at 4~kPa and \textbf{c}, \textbf{d}, and \textbf{f} at 5~kPa. Pictures \textbf{b} and \textbf{d} were captured with the 550-nm filter. A uniform 3.56-mm-wide test section was used for all pictures except \textbf{e} and \textbf{f} where the test section was intentionally canted. The 254-mm-diameter insert was used for all the pictures shown.}
    \label{fig: open-shutter}
\end{figure*}

Several open-shutter pictures of the imploding cylindrical detonations were taken to capture the cellular structure. Figure~\ref{fig: open-shutter} shows pictures of the stoichiometric acetylene-oxygen detonations under various conditions. Both the support airfoils and the 254-mm-diameter wave-shaping insert were used. The window was partially covered with an opaque ring used to reduce the viewing area to approximately 180~mm in diameter. This was done to block the light emitted by the detonation wave before entering the thin test section, which would otherwise make capturing the much dimmer cellular detonation in the test section difficult. Initial pressure before initiation was set to 2.5~kPa for pictures \textbf{a} and \textbf{b}, set to 4~kPa for picture \textbf{e} and set to 5~kPa for pictures \textbf{c}, \textbf{d}, and \textbf{f}. The difference between pictures \textbf{a} and \textbf{b}, and \textbf{c} and \textbf{d} is that a 550-nm short-pass filter was added to the camera for pictures \textbf{b} and \textbf{d}. Pictures \textbf{e} and \textbf{f} illustrate the shape of the detonation structure after the test section was intentionally canted (see details given in section \ref{sec422}). All pictures, including those with the same initial pressure, were taken in different tests. The pictures were post-processed to enhance cellular structure visibility by modifying the texture and final exposure of the images. 

As seen in the detonations initiated at 2.5~kPa, the lopsidedness of the cellular structure is indicative of a detonation near failure, where different sections of the front may have reverted to a deflagration before reaching the test section. In contrast, detonations initiated at higher pressure display a more uniform pattern. For detonations in a canted test section, cell size appears larger on the side closer to the implosion focus. Also captured in the pictures are combustion by-products, which appear orange, presumably due to continuous (black body) emission from condensed phase products.

\section{High-speed videography}\label{sec4}

\begin{figure*}[ht]
    \centering
    \includegraphics[width=0.75\textwidth]{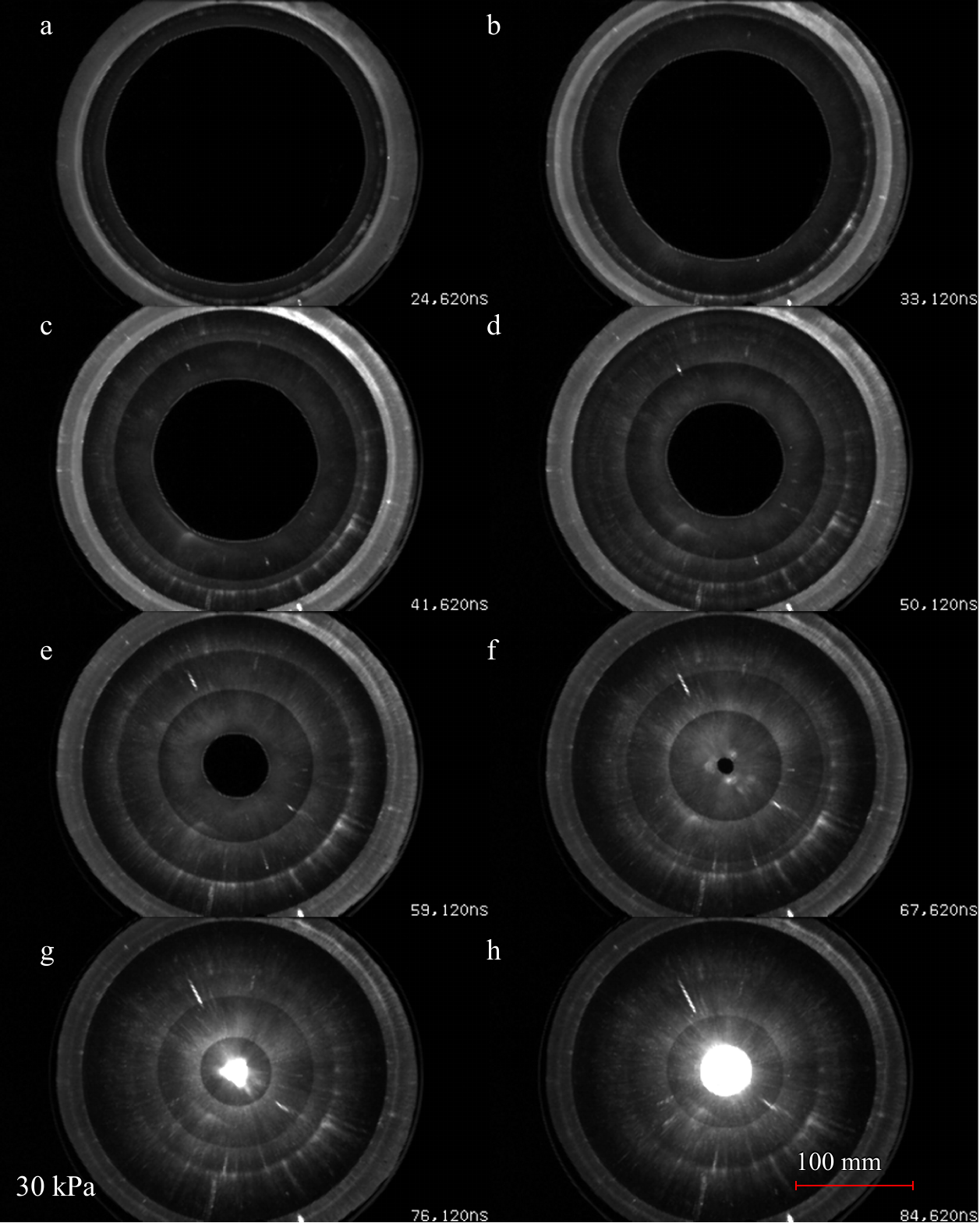}
     \caption{Frames showing a converging detonation as recorded from the high-speed video. The detonation wave was initiated at 30~kPa and the 254-mm-diameter insert was set to a constant 3.56-mm channel width. Concentric circles following the converging front are shock waves created at the moment the detonation enters the cavity and impacts the center disk. High-speed videos of the converging detonations are found in the supplementary material.}
    \label{fig: video frames}
\end{figure*}

The following are the results collected for this experiment through high-speed videography. The Shimadzu HPV-X2 camera was set to 2,000,000~FPS, 200~ns of shutter time, and a varied aperture to ensure optimal visualization of the wavefront. Figure~\ref{fig: video frames} shows an array of frames displaying a converging detonation as captured in the apparatus. This example corresponds to a detonation initiated at 30~kPa initial pressure, and with the large wave-shaping insert (254~mm diameter) in place, limiting the channel width in the test section to 3.56~mm. The detonation is seen immediately after it arrives in the test section and converges to the center.

\subsection{Image processing from high-speed videos}\label{section: data processing}

Several parameters were collected from the video of each test to characterize the behavior of the imploding detonations. A MATLAB script employing the image processing toolbox \cite{MATLAB} was developed to analyze the videos. This script worked by first separating all frames showing the detonation front within the test section for individual analysis. Then, an edge-finding algorithm was used to approximate the perimeter, area, and centroid of the shape bound by the detonation front as it converged. This data was subsequently used to determine three geometric parameters of interest. Also, the velocity at different front locations was estimated; the process details are described below. 

The three geometric parameters used to characterize the imploding detonations are the symmetry ratio, the off-center offset, and the azimuth. Because the symmetry ratio cannot be measured at the center of convergence, the trend in which these parameters develop over the frames studied was used to extrapolate the parameters to their value at the convergence center through a linear fit.  

\begin{figure}[!t]
\centering
\includegraphics[width=\columnwidth]{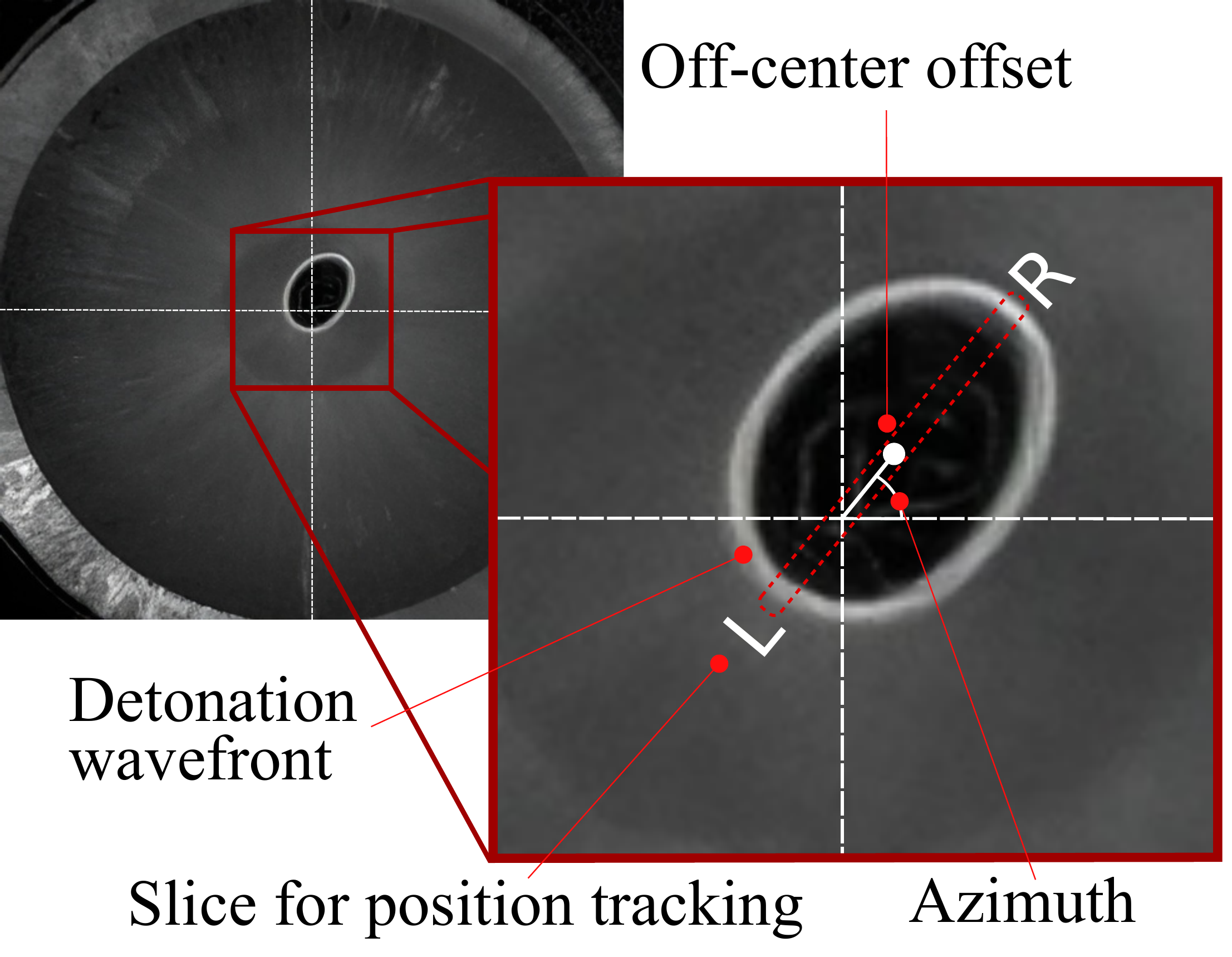}
\caption{Diagram showing the geometric parameters extracted from high-speed videos of frames before convergence. The slice used to measure the velocity of the front at the left and right sides of the center of convergence is also shown. Slices taken were always aligned with the measured azimuth.}
\label{fig: data analysis}
\end{figure}

The symmetry ratio is defined as

\begin{equation}
    \frac{4\pi A}{p^2}
    \label{equation: ratioasym}
\end{equation}
which approaches unity as the front approximates a circle. In the equation, $A$ corresponds to the area bounded by the front, and $p$ is the perimeter of the area. The off-center offset of the detonation was measured as the distance between the centroid of the front and the center of the test section, which was obtained from a calibration picture taken for each test batch. The azimuth of the detonation center was calculated to determine its angular position with respect to the center of the test section. Figure~\ref{fig: data analysis} illustrates the different geometric parameters extracted from the video frames. Note that the primary source of uncertainty in the measurement originates from the limited image resolution coupled with the self-luminous videos used. Based on the video calibration used, it is expected for the wavefront position uncertainty to be bound to approximately $\pm 1~$mm.

The pressure range for all high-speed videography tests is 5 to 40~kPa. No tests done at 2.5~kPa were included in the results because, at this initial pressure, local extinction of the detonation front was frequent, which led to poor reproducibility. Experimental cell size data for the acetylene-oxygen mixture was obtained from \cite{Manzhalei,desbordesCriticalDiameterDiffraction1985,desbordesTransmissionOverdrivenPlane1987} and fitted to approximate the cell size over the experimental pressure range. The relation used is
\begin{equation}
\lambda (P)= 28.66 \cdot P^{-1.139}
\end{equation}
where pressure, $P$, is in kilopascals and cell size in millimeters. All the data displayed in the results section is plotted over the width of the test section $w$, normalized by the cell size $\lambda$. In cases where the test section width is not uniform, the width at the center of the apparatus was used.

To capture the velocity of the front, a one-dimensional slice of a fixed section of the view area was taken from all frames sampled (as shown in Fig.~\ref{fig: data analysis}). These slices were then stacked in an array where the columns correspond to the different frames, and the rows are the slices showing the detonation. The front location for each frame was determined through a finding algorithm that probed pixel by pixel until it found the front based on changes in light intensity. Then, the velocity at the front was approximated by fitting a curve through the pixels showing the detonation front. The velocity reported in the results corresponds to the average velocity before convergence. High-speed videos would often be rotated so that the sampled sections of the detonation front were aligned with the azimuth, and off-center offset data was used to select the location of the slices, always matching opposing sides of the converging wavefront. 

\subsection{Symmetry dependence on channel width} \label{section: symmetry data}
As mentioned in section \ref{section: data processing}, several parameters were collected from the different high-speed videos of the imploding detonations under varying initial pressure. For these tests, the geometry of the test section was characterized by measuring the channel width throughout the test section with a micrometer, as described in section \ref{sec2}. The tests performed are divided into \emph{constant-width} and \emph{varying-width} tests. 

\subsubsection{Constant-width tests}

\begin{figure}
    \begin{subfigure}[b]{\columnwidth}
        \includegraphics[width=\linewidth]{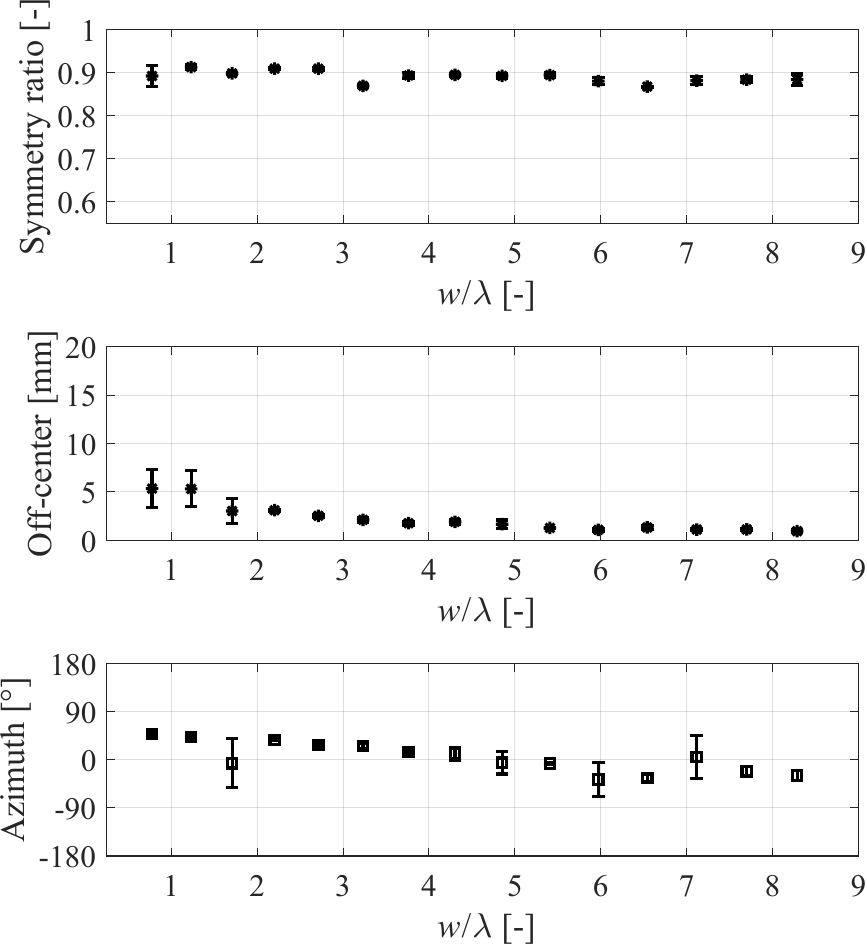}
        \caption{Geometric parameters extracted from high-speed videos.}
        \label{fig: constant-width_a}
    \end{subfigure}
    \hfill
    \begin{subfigure}[b]{\columnwidth}
        \includegraphics[width=\linewidth]{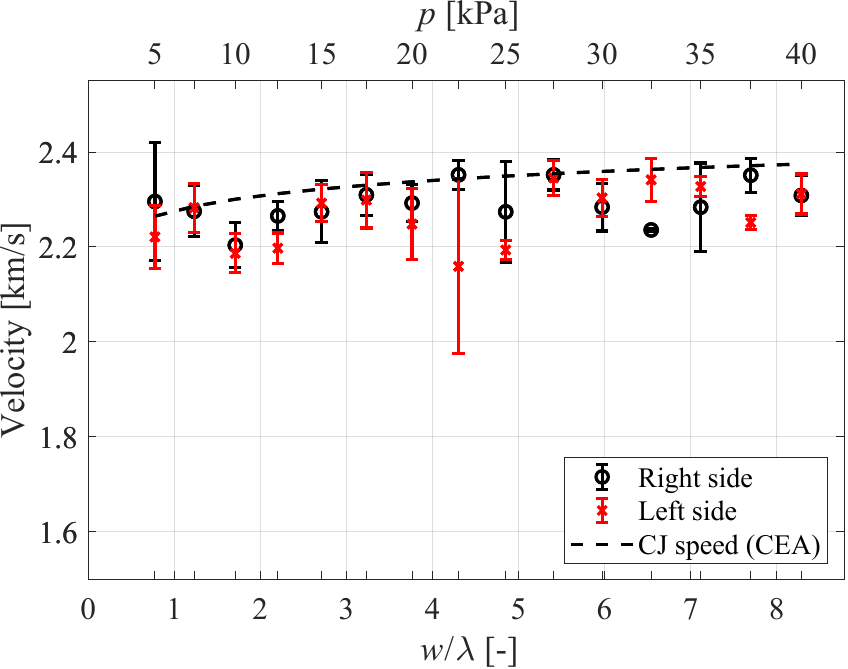}
        \caption{Measured front velocity at opposing sides of the front. The experimental data is compared to the expected CJ speed.}
        \label{fig: constant-width_b}
    \end{subfigure}
    \caption{Measured data from tests with constant channel width. Tests were done with the 204-mm-diameter insert with a uniform 3.56-mm channel width. The pressure range is 5 to 40~kPa in increments of 2.5~kPa. Error bars represent the standard deviation between the multiple tests done at each initial pressure. The width at the center of the test section was used to calculate $w/\lambda$ in all tests.}
 \label{fig: constant-width}
\end{figure}

This corresponds to the first series of tests done with the apparatus, where the test section was set to have a constant channel width of 3.56~mm with a measured variation of less than 6$\%$, which was attributed to the machining tolerances of the apparatus. Figure~\ref{fig: constant-width} shows the data collected from tests with initial pressure between 5 to 40~kPa. Each data point represents the average of at least three tests, and error bars are the calculated standard deviation for each set. Figure~\ref{fig: constant-width_a} contains the geometric parameters that were extracted from the high-speed videos, and \ref{fig: constant-width_b} shows the measured front velocity at two opposing sides on the front, which are aligned to the azimuth calculated for each test. The CJ velocity of stoichiometric acetylene-oxygen at room temperature for the same pressure range was estimated using the NASA CEA code \cite{CEA} and is included in the figure.

As seen in Fig.~\ref{fig: constant-width_a}, all constant-width tests display a high degree of symmetry, evidenced by a consistent symmetry ratio of around 0.9 and a maximum off-center offset of only 5~mm, compared to the overall diameter of the center disk of 330~mm. The off-center offset also increased as the number of cells across the width of the channel approached unity ($w/\lambda = 1$). The azimuth determined varies significantly between $-30^\circ$ to $30^\circ$, which is to be expected considering the low off-center offset measured. At the same time, the detonation velocity measured at opposing sides of the converging detonations was found to be close to the expected CJ velocity, and no increase in velocity of the wave as it converged toward the center was noted (see section \ref{sec5} for further details). 

\subsubsection{Varying-width tests} \label{sec422}
\begin{figure}[ht]
    \centering
    \includegraphics[width=\columnwidth]{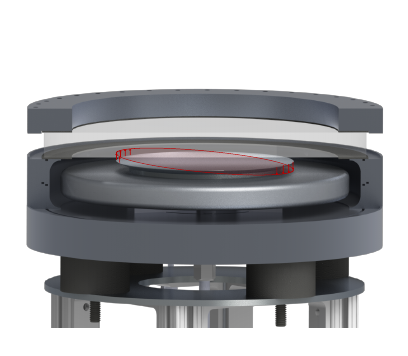}
    \caption{Demonstration of how the insert is canted for experiments with varying test section width. The canting was exaggerated in this image for better visualization.}
    \label{fig: canting render}
\end{figure}

Through experience with the apparatus, the experiment was found to be extraordinarily sensitive to the way it was assembled and that small variations in the setup, like uneven compression of the rubber seals, would lead to inconsistencies in the results, evidenced by an off-center offset in the center of the implosion. This finding motivated the deliberate canting of the wave-shaping insert to quantify the influence of the nonuniformity in the width of the test section on the results. For these tests, the test section was intentionally canted by inserting precisely measured metal shims under the wave-shaping insert. Figure~\ref{fig: canting render} gives an exaggerated illustration of how the test section shape changed after canting, and Table~\ref{tab: 1} reports the canting angle, which was calculated from measured channel width data for each of the tests with different wave-shaping inserts. 

\begin{figure*}[!h]
    \centering
    \begin{subfigure}[b]{0.49\textwidth}
      \centering
      \includegraphics[width=\linewidth]{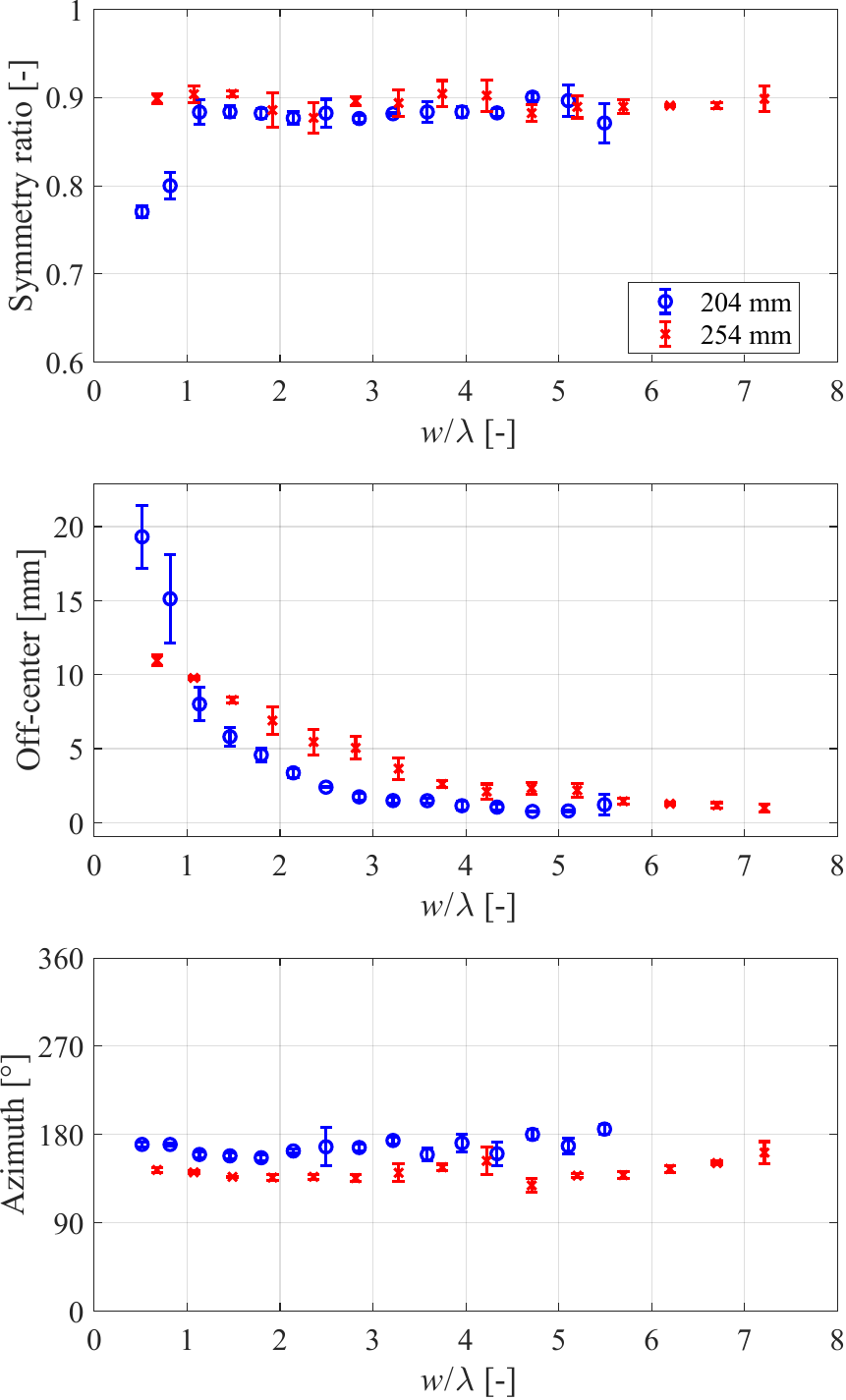}
      \caption{Geometric parameters extracted from high-speed videos. The two data sets correspond to tests done with the small (blue) and large (red) inserts.}
      \label{fig: canting_a}
     
    \end{subfigure}%
    \hfill
    \begin{subfigure}[b]{\columnwidth}
      \centering
      \begin{subfigure}{\linewidth}
        \centering
        \includegraphics[width=\linewidth]{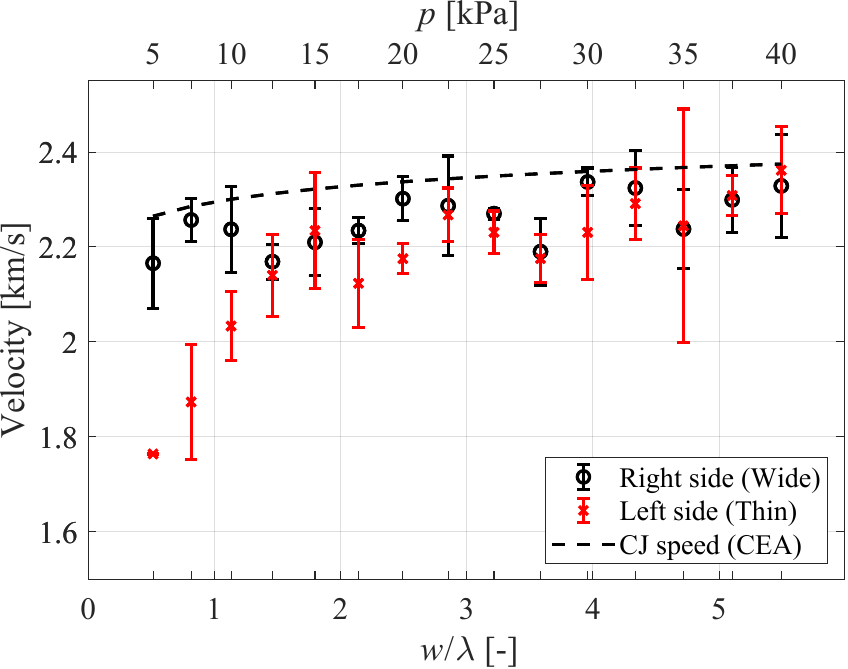}
        \caption{Measured front velocity for the small wave-shaping insert tests. The experimental data is compared to the expected CJ speed.}
          \label{fig: canting_b}
      \end{subfigure}
    
      \begin{subfigure}{\linewidth}
        \centering
        \includegraphics[width=\linewidth]{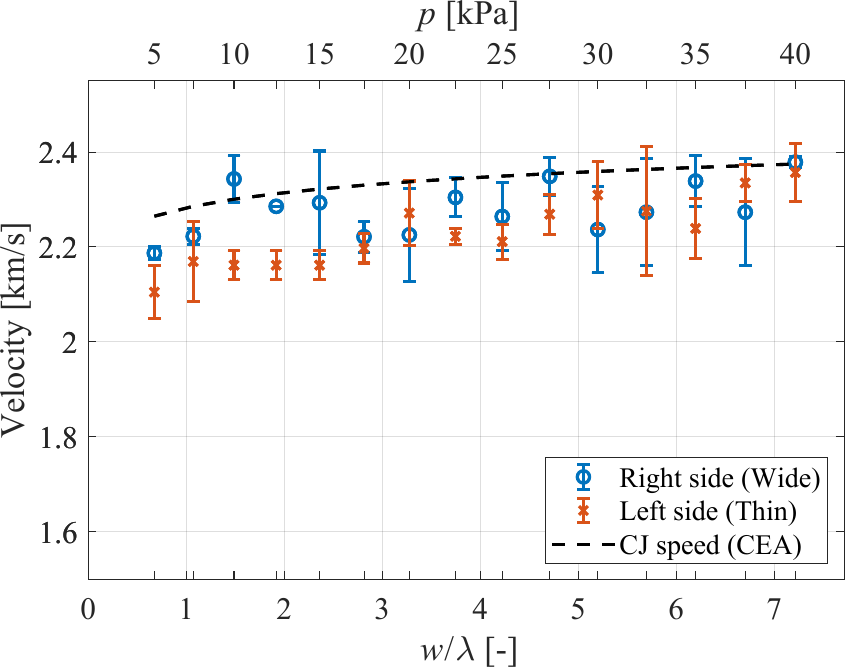}
        \caption{Measured front velocity for the large wave-shaping insert tests. The experimental data is compared to the expected CJ speed.}
          \label{fig: canting_c}
      \end{subfigure}
    \end{subfigure}
    
    \caption{Measured data for varying-width tests. Tests were done with the 204-mm and 254-mm-diameter inserts, with the canting angle described in Table~\ref{tab: 1}. Error bars represent the standard deviation between the multiple tests done at each initial pressure. The width at the center of the test section was used to calculate $w/\lambda$ in all tests.}
      \label{fig: canting}
  \end{figure*}
  
\begin{table}[ht]
\caption{Canting information for each of the test runs with different insert sizes. The angle was defined from the side of least width to the side of maximum width. The test section width was measured at the center of the wave-shaping insert.}
\label{tab: 1}
 \begin{tabular}{c|cc}
        \toprule
        \textbf{Insert} & \textbf{Canting angle [$^\circ$]} & \textbf{$w_{\mathrm{center}}$ [mm]} \\ \midrule
        \textbf{204 mm}              & 0.63                              &2.36                              \\
        \textbf{254 mm}              & 0.34                              & 3.10                               \\ \bottomrule
    \end{tabular}
\end{table}

Figure~\ref{fig: canting} displays the data collected for the varying-width tests with both the large and small inserts. The geometric parameters shown in Fig.~\ref{fig: canting_a} present a contrasting behavior to what is seen in the constant-width tests. The symmetry ratio is close to unity and is consistent for the large insert but decreases with a reduced number of cells per channel width for the small insert. In the case of the off-center offset, both datasets have a trend where the off-center offset grows significantly as the number of cells diminishes, reaching a maximum value of about 10~mm for the large insert and 20~mm for the small one. Last, contrary to the constant-width tests, both data sets have a uniform and distinct azimuth. The azimuth detected in both cases matches the location where the channel width was the thinnest or where the metal shims were placed. 

In the case of velocity, as shown in Figs. \ref{fig: canting_b} and \ref{fig: canting_c}, a trend similar to that of the off-center offset is seen in the velocity measured on the thinnest side of the front. Here, a deficit from the CJ velocity grows as the number of cells within the channel decreases. This trend is present in both small and large insert tests. The fact that the large insert tests show a lesser asymmetry is believed to come from the fact that the large insert was not canted as much as the smaller one. This was done to prevent completely closing one side of the test section with the edge of the larger insert.


\section{Analysis} \label{sec5}
\begin{figure}[t]
    \centering
    \includegraphics[width=\columnwidth]{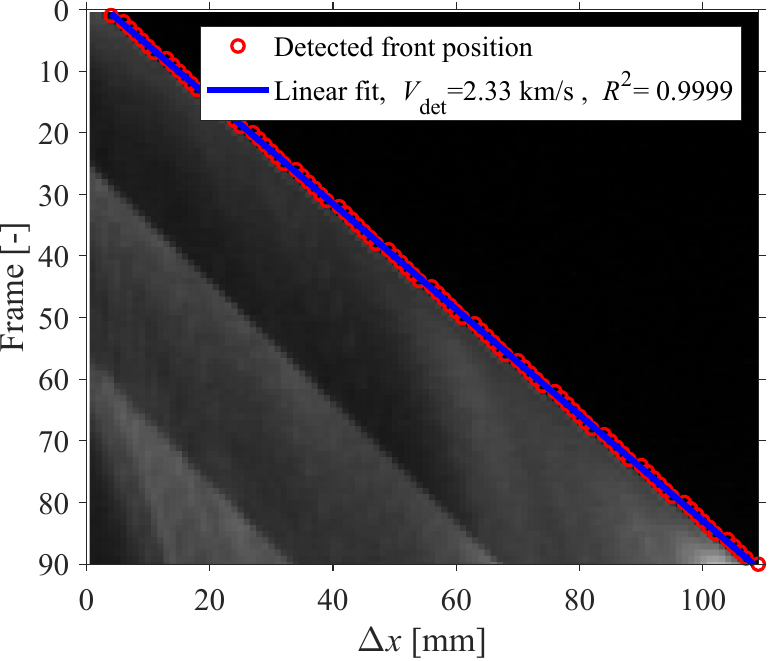}
    \caption{Space--time ($x$--$t$) plot of a 30 kPa detonation in a constant-width section. Lines extracted from video frames are stacked vertically, showing detonation entry in the test section (upper left) and propagation to the apparatus center (lower right). The large wave-shaping insert was used.}
    \label{fig: velocity tracking}
\end{figure}

As shown in section \ref{section: symmetry data}, variations in the width of the test section affect the symmetry of the converging fronts, evidenced by the significant off-center offset and consistent azimuth measured. Two hypotheses were proposed to explain the observed behavior:
\begin{enumerate}
\item The velocity deficit is caused by the change in the local front area relative to the area of the front at the start of convergence, a phenomenon often described using the CCW method for geometric shock dynamics. The variation in width across the channel would result in an asymmetry in the degree of area convergence. \\
\item The velocity deficit is caused by a phenomenon described by Fay in \cite{fayTwoDimensionalGaseousDetonations1959} where the detonation wavefront loses momentum when confined within a thin (width on the order of a few cells) channel because of the negative displacement of the boundary layer, with the extent of the velocity deficit depending on the relative scale between the cell size and the width of the channel.
\end{enumerate}
It is possible to rule out hypothesis (1) by recognizing that---although some thermodynamic properties like pressure vary significantly through area convergence---detonation velocity is mainly insensitive to this effect, a phenomenon experimentally verified by \cite{ahlborn}. Through their experiments, the detonation velocity was found to only increase by about 20\% after the radius of convergence had reached a twentieth of its initial size. This finding was also verified in the present study through the velocity estimated from high-speed videos. Figure~\ref{fig: velocity tracking} shows the curve used to estimate the velocity of an imploding detonation in a constant-width test section. Here, the red circles correspond to the tracked position of the detonation front at each frame, and the blue line is a linear fit of this data. The goodness of the fit was characterized by the $R^2$, and the estimated detonation velocity is included in the figure. As seen in the figure, there is no noticeable velocity deficit caused by the decrease in the radius of convergence of the front as the detonation converges. Under the same principle, the change in the local front area (which varies much less than the radius of convergence) from propagating through a varying-width channel is unlikely to cause the off-center offset observed in the results. In addition, if area convergence because of the canting was the cause of the offset, the effect should be independent of pressure. However, the data shows an increase in the off-center offset with decreasing initial pressure (increasing detonation cell size).

To confirm whether boundary layer displacement---Hypothesis (2) above---is the mechanism behind detonation wave asymmetry, a numerical model based on a Huygens construction was formulated to simulate the dynamics of a converging detonation in a thin channel of nonuniform width. In this model, the detonation wavefront is modeled as a series of small planar fronts that converge based on the orientation of neighboring fronts and a locally constant velocity set by the channel width. This approach is a treatment similar to that done by Neemeh in \cite{Neemeh}, who studied cylindrical shock convergence in narrow channels based on the formulation by Mirels \cite{Mirels}.

\subsection{Wavefront initiation and propagation}

This detonation front was initialized in a cartesian plane as $N$ points around a perfect circle with a diameter equal to the diameter of the test section. The front was thus divided into $N-1$ local fronts, each defined as the line connecting two neighboring points. Here, the position of the first point at the first step is given by
\begin{equation}
   \mathbf{p}^{j}_{i}=\mathbf{p}^{j}_{i}(x,y)
\end{equation}
with $j$ defining the current time step, and $i$ defining the current point. A study of grid independence for this model was performed, and convergence of results was found if at least 100 elements were used. 

\begin{figure}[t]
    \centering
    \includegraphics[width=\columnwidth]{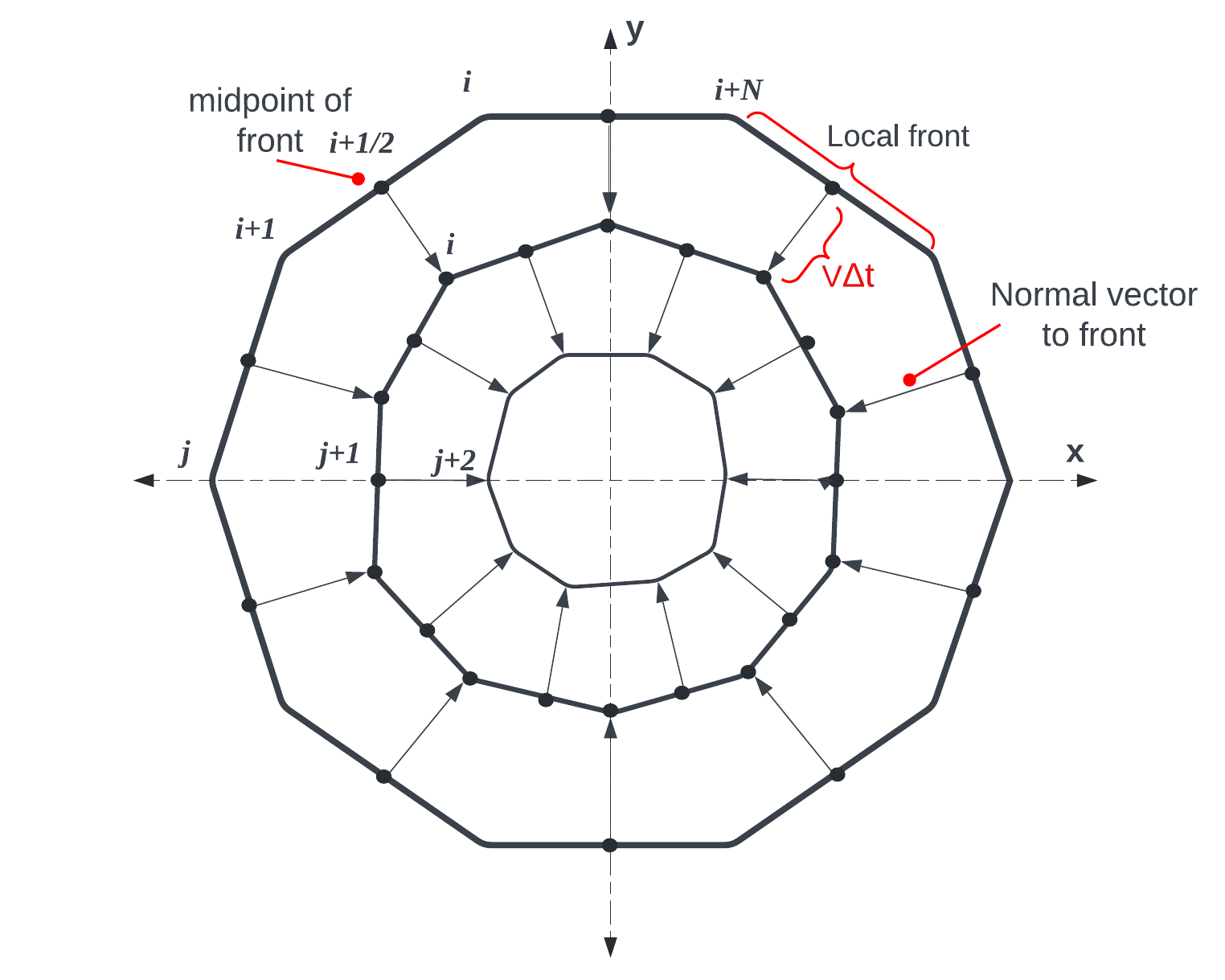}
    \caption{Schematic of the numerical model.}
    \label{fig: num model}
\end{figure}

The propagation of the converging front was done by first tracing a line connecting points $i$ and $i+1$. Then, the inwards-pointing normal vector of the line was determined and used as the direction in which this local front propagated. The point $i+1/2$, the midpoint between $i$ and $i+1$, was then found and used as the position from which to propagate the local front forward. After the direction and origin point were determined for all of the small fronts, each was propagated inwards as 
\begin{equation}
     \mathbf{p}^{j+1}_{i} =  \mathbf{p}^{j}_{i+1/2}+\mathbf{v}^{j}_{i+1/2} \mathrm{\Delta} t
        \label{equation:pos_eq}
\end{equation}
with $\mathbf{v}^j_{i+1/2}$ the local velocity, and $\mathrm{\Delta }t$ a fixed time step. Subsequently, point $i+1/2$ became point $i$ for the new time step, and the process was repeated. Figure~\ref{fig: num model} is a graphical representation of this model. 

\subsection{Local velocity estimation}
The local velocity $\mathbf{v}^j_{i+1/2}$ was determined by implementing the velocity deficit through the boundary layer displacement theory described in \cite{fayTwoDimensionalGaseousDetonations1959}. This approach approximates the detonation as quasi-one-dimensional (i.e., cellular structure is not considered) and adiabatic. The boundary layer displacement thickness is thus given by 
\begin{equation}
    \delta^*=0.22\,t^{0.8} \left ( \frac{\mu_\mathrm{e}}{\rho_1 v}  \right)^{0.2}
    \label{equation:disp_thick}
\end{equation}
where $\rho_1$ is the upstream density, $\mu_\mathrm{e}$ the viscosity of the gas in the reaction zone, $t$ is the hydrodynamic thickness of the detonation, defined as $1.5\lambda$, and $v$ is the detonation velocity. Transport properties of the stoichiometric mixture at room temperature were obtained using the NASA CEA program. 

The velocity deficit is given by
\begin{equation}
    \frac{v}{v_{\mathrm{CJ}}}=1-0.53\,\epsilon\,\zeta
    \label{equation: vel_def}
\end{equation}
where $\zeta$ is the fractional increase in the area of any stream tube compared with its area at the shock front, and $\epsilon$ is a parameter that relates to the pressure gradient across the detonation front. The expression for $\zeta$ is given by
\begin{equation}
    \zeta=\frac{2\delta^* }{w}
    \label{equation: zeta}
\end{equation} 
where $w$ is the local channel width, obtained from measurements as described in section \ref{sec2}. For the numerical simulations, the channel width at the position of each midpoint, $\mathbf{p}^j_{i+1/2}$, was used as the width of the local front at that location. 

The parameter $\epsilon$ is a constant that depends on the pressure rise across the reaction zone, which is expected to be between 1 and 2 \cite{fayTwoDimensionalGaseousDetonations1959}. This constant was determined for this experiment by comparing the velocity deficit model with experimental data of velocity deficits of cylindrical detonations in stoichiometric acetylene-oxygen obtained by Ng et al. \cite{Ng} in a setup similar to the current geometry (thin cylindrical channels). In this case, $\epsilon\approx~1.4$ was found to match the experimental data well. In summary, the local velocity was determined from equations \ref{equation:disp_thick}, \ref{equation: vel_def}, and \ref{equation: zeta} using a root-finding method (fzero function in MATLAB) to determine the local velocity of the front. Alternatively, the CJ speed may be used to estimate the boundary layer displacement thickness $\delta^*$, in which case root-finding is not needed. Both methods were tested and resulted in similar trends, although the use of the CJ speed will result in an under-prediction of $\delta^*$.

\section{Discussion} \label{sec6}
As mentioned in section \ref{sec5}, based on the results obtained from the varying-width tests, it is expected that the asymmetry of the imploding detonations is produced by a local velocity deficit caused by the negative displacement of the boundary layer, and manifested by a low number of cells available within the test section. 

\begin{figure} [ht]

    \begin{subfigure}[b]{\columnwidth}
    \centering
    \includegraphics[width=\linewidth]{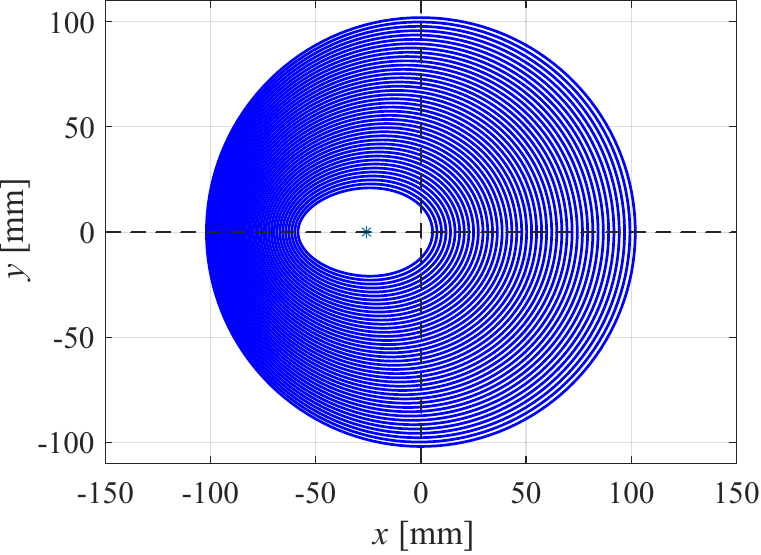}
    \caption{Output from numerical simulation, based on width data of the 204-mm-diameter insert tests and 5~kPa initial pressure. $N=150$ and $\mathrm{\Delta} t=1.5~\upmu$s.}
    \end{subfigure}

    \begin{subfigure}[b]{\columnwidth}
    \centering
    \includegraphics[width=0.95\linewidth]{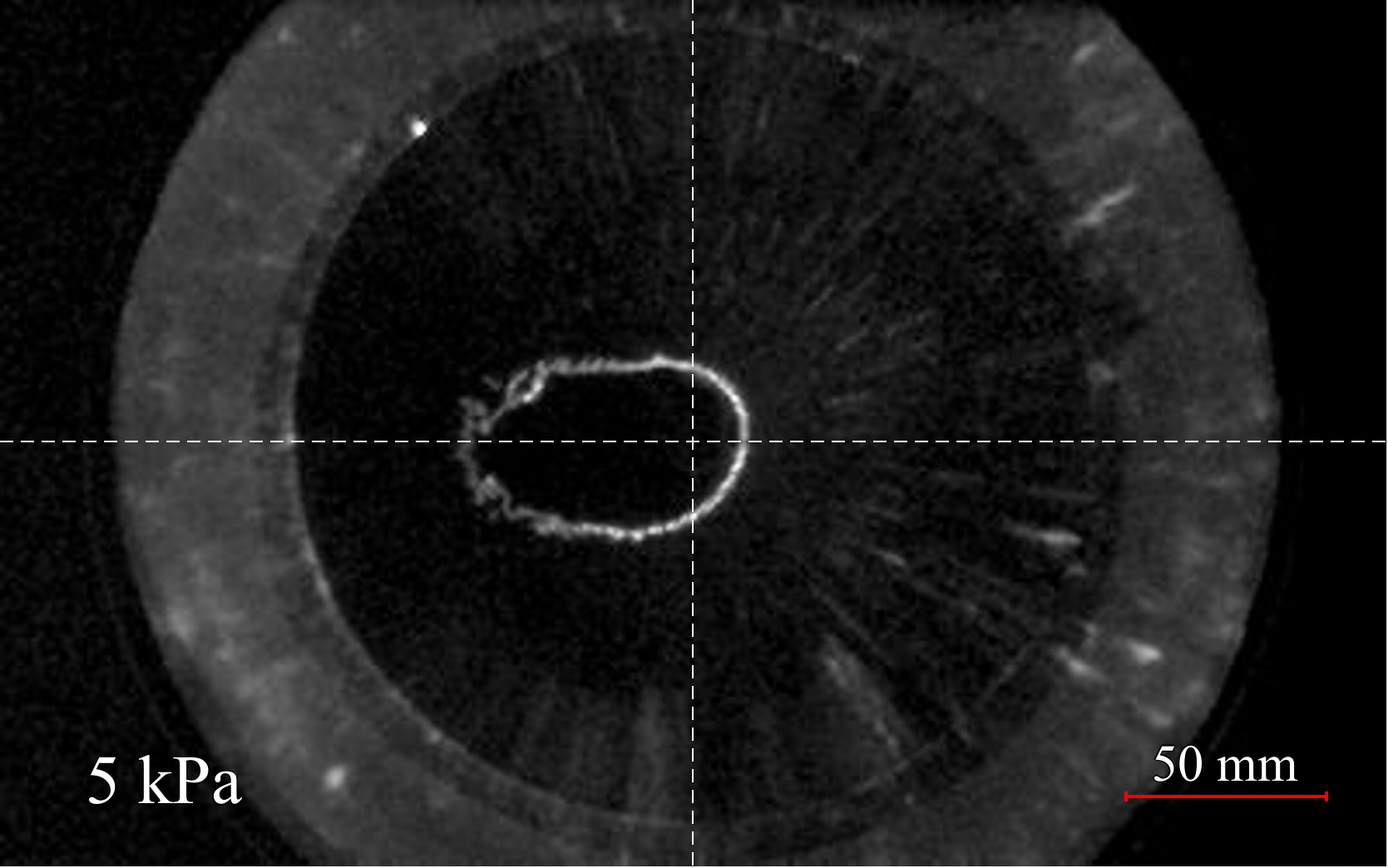}
    \caption{Frame from a high-speed video showing imploding detonation with 5~kPa initial pressure using the 204-mm-diameter insert.}
    \end{subfigure}
    \caption{Comparison between the output from the numerical model and a high-speed video test. For both the experiment and model, the thinnest side is located at the left (corresponding to an azimuth of 180$^\circ$).}
    \label{fig: comparison qualitative}
\end{figure}

\begin{figure}[t]
    \centering
    \includegraphics[width=\columnwidth]{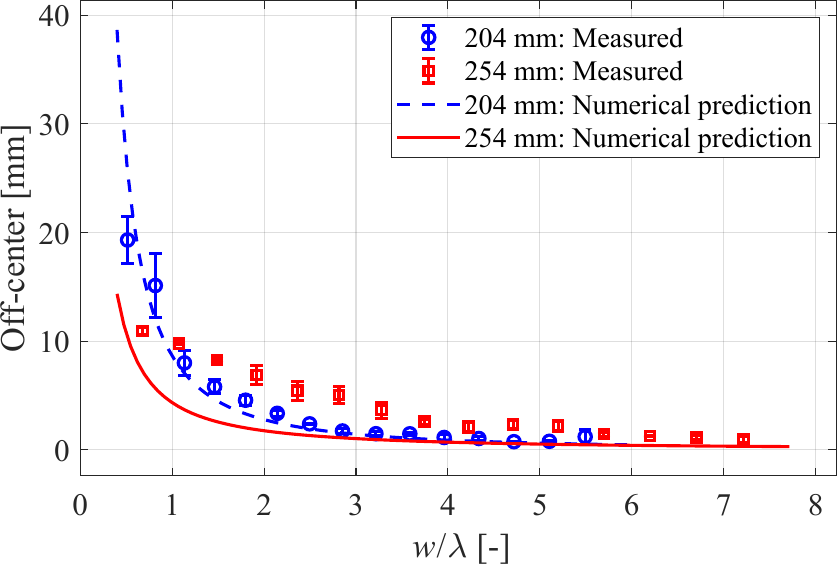}
    \caption{Comparison between experimental results and numerical predictions of the varying-width tests.}
    \label{fig: comparison numerical}
\end{figure}

In such a case, the numerical model formulated, which depends only on the conditions of the test section and gas, should match the measured off-center offset from the high-speed videos. Figure~\ref{fig: comparison qualitative} is a comparison between the numerical model output and a high-speed video from the tests using the 204-mm-diameter canted insert, both at 5~kPa initial pressure. Also, Fig.~\ref{fig: comparison numerical} compares the off-center offset measured for the different tests and the one predicted with the numerical model, using the test section data from Table~\ref{tab: 1} to determine local front velocity.

As seen in Fig.~\ref{fig: comparison qualitative}, the numerical model adequately reproduces the features observed in the high-speed video, both matching the axis in which the front is asymmetric and the shape in which this asymmetry is present. Features that the numerical model does not capture are the presence of striations and irregularities that are seen on the detonation front, which form as the detonation weakens by propagating through the thin channel and cell size increases. The fact that the model does not reproduce these features is expected since this model does not account for transverse waves at the detonation front. The implosion of this oval-shaped front along a line (rather than the point-implosion of a symmetric detonation) is believed to give rise to the line of intense luminosity observed at the center of convergence in open-shutter photographs of asymmetric detonations, as seen in picture \textbf{f} from Fig.~\ref{fig: open-shutter}.

Figure~\ref{fig: comparison numerical} shows that the model effectively reproduces the behavior of increasing off-center offset with a decreasing number of cells within the test section. Particularly, it adequately recreates the off-center offset measured from both test sets shown. Note that for the tests using the 254-mm-diameter insert, there is some discrepancy between the measurements and the model curve. It is possible that because the insert is wider, disturbances of the wavefront generated as the detonation turns around the center disk increased, thus explaining the inconsistency in the results. It is also expected that other discrepancies between the numerical model and results come from secondary phenomena not accounted for by the model, such as the presence of a cellular structure, or the limitations of the measurement process, which is constrained by image resolution and the accuracy with which the width of the test section can be determined. The wavefront-tracking limitations may be partially remediated using schlieren visualization instead of self-luminous videography. 

In any case, as evidenced by the adequate agreement between the model and the data, it is confirmed that the primary mechanism that causes asymmetry of the detonations is the velocity deficit due to boundary layer displacement. In the present experiment, the non-uniformity in channel width is what led to asymmetry in the converging front.

\section{Conclusions}
The use of a cylindrical test section with a width on the order of the detonation cell size (i.e., a thin channel) has permitted the dynamics of imploding detonation waves in acetylene-oxygen to be observed in a two-dimensional geometry using both open-shutter photography and high-speed videography. While this technique permits the three-dimensional dynamics of the wave to be suppressed, it results in an additional complication in that the wave dynamics become sensitive to the assembly of the apparatus (e.g., nonuniform compression of the rubber seals in assembling the apparatus may result in observable asymmetry and off-set from the apparatus center as the detonation implodes). The detonation wave also exhibited a remarkable persistence of perturbations that were introduced to the wave as it entered the apparatus and transitioned to a cylindrical detonation. To further investigate the mechanism responsible for the asymmetry of the implosions, the width of the test section was intentionally canted to result in a slight gradient in channel width (0.3--0.6$^\circ$) across the test section. A Huygens-like wave propagation model, wherein the local width of the channel determines the instantaneous propagation velocity of the detonation, was able to reproduce the asymmetries and off-sets observed in the experiments successfully. This agreement between the experiment and model confirms that momentum losses to the walls of the thin channel are the primary cause of the asymmetry observed in this investigation. The findings of this study can guide future use of this (and similar) apparatus in investigating the states that may be obtained in imploding detonations.

\section*{Acknowledgments}
The authors wish to thank Melody Haerens, Fariha Hassan, Jaden Cheng, and Pedro Costa Ferreira Pereira Leite for the design of the experimental apparatus and Andreas Hofmann, Ramnarine Harihar, and Meisam Aghajani for the fabrication and technical support in modifying the apparatus. Simeon Radev is acknowledged for the design of the airfoil supports. The Canadian Armed Forces Joint Counter Explosive Threat is thanked for the use of the Shimadzu HPV-X2 camera. The authors would also like to acknowledge helpful discussions with Charles Kiyanda and Ashwin Chinnayya. We also thank the reviewers for their helpful feedback. This research was supported by the Natural Sciences and Engineering Research Council of Canada and the McGill Summer Undergraduate Research in Engineering program. The authors declare no conflict of interest.

\section*{Data availability}
The authors declare that data supporting the findings of this study are available within the article. High-speed video and open-shutter image examples are found within the supplementary material. Should any raw data files be needed in another format they are available from the corresponding author upon reasonable request.

\section*{Appendix: Influence of the support structure}
\begin{figure}[h]
    \begin{subfigure}[b]{0.99\columnwidth}
        \centering
        \includegraphics[width=\linewidth]{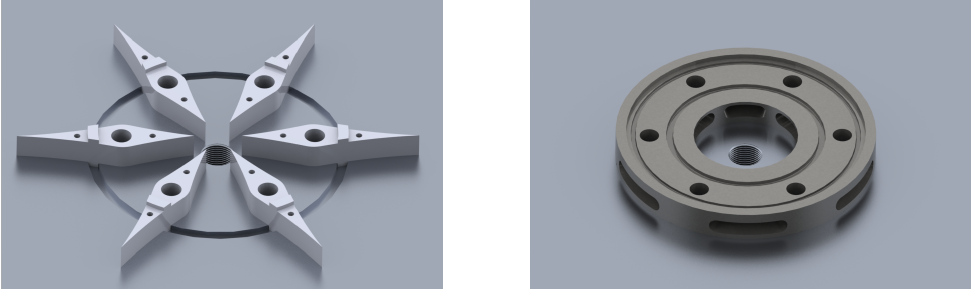}
        \caption{Renders of support structures for the center disk: airfoils (left), cylinder (right), with detonation inlet at the center.}
        \label{fig: supports}
    \end{subfigure}
        
    \begin{subfigure}[b]{0.99\columnwidth}
        
        \centering
        \includegraphics[width=\linewidth]{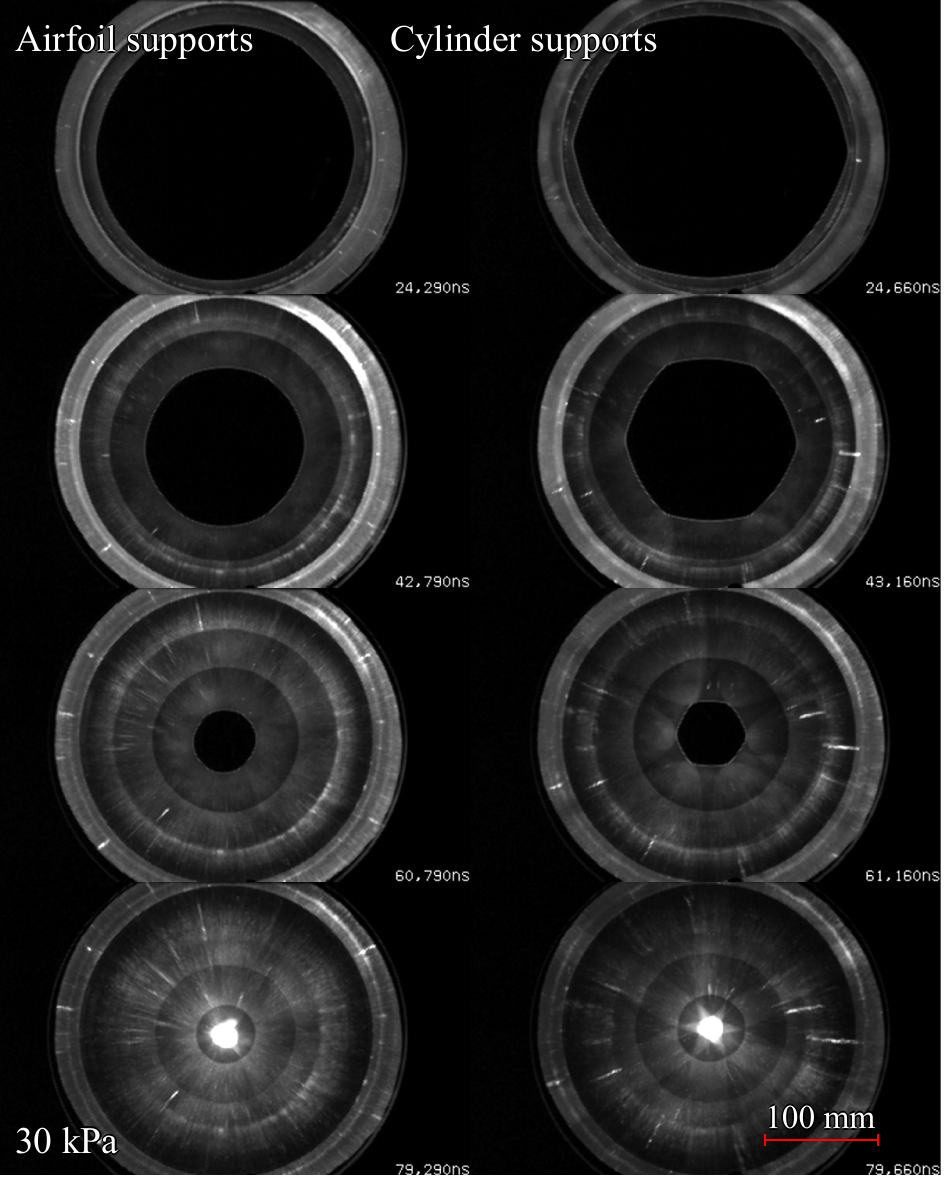}
        \caption{Comparison between tests with different support structure versions.}
        \label{fig: support comparison}
    
        \end{subfigure}
        \caption{Influence of the support structure on the shape of the imploding detonations. The 254-mm-diameter insert was used for this test and the test section width was set to a uniform 3.56~mm.}  
\end{figure}

Two different designs for the supports of the center disk were tested, as shown in Fig.~\ref{fig: supports}. The first is a cylinder with six equally spaced diverging channels, and the second consists of six individual diamond-shaped airfoils with extended trailing edges. Figure~\ref{fig: support comparison} compares tests done with each of the two configurations. 

From comparing the tests done with the cylinder and airfoils, it is clear that the support structure exerts a significant influence on the final symmetry of the front, even with the curvature distribution mechanism mentioned in \cite{Lee_3}. As seen in Fig.~\ref{fig: support comparison}, the cylinder support imprints a hexagonal pattern onto the converging front, which is visible until the detonation converges. This phenomenon is explained by the interaction of different detonation waves exiting the cylinder and producing Mach stems as they come in contact. In contrast, this effect was minimized with the airfoils, allowing for a smooth imploding front to be formed. Note that an investigation on the impact of diamond-shaped airfoils in the path of propagating planar detonation waves was done in \cite{knystautasDiagnosticExperimentsConverging1969}.

\bibliography{bibliography}

\begin{thebibliography}{10}
\providecommand{\url}[1]{{#1}}
\providecommand{\urlprefix}{URL }
\providecommand{\doi}[1]{\url{https://doi.org/#1}}
\bibcommenthead

\bibitem{Guderley1942}
G.~Guderley, Starke kugelige un zylindrische verdichtungsstobe in der nahe des kugelmittelpunktes bzw der zylinderachse.
\newblock Luftfahrtforschung \textbf{19}(302) (1942)

\bibitem{Gardner1982}
J.H. Gardner, D.L. Book, I.B. Bernstein, Stability of imploding shocks in the ccw approximation.
\newblock Journal of Fluid Mechanics \textbf{114}, 41--58 (1982).
\newblock \doi{10.1017/S0022112082000032}

\bibitem{jacksonPDE}
S.I. Jackson, J.E. Shepherd, Toroidal imploding detonation wave initiator for pulse detonation engines.
\newblock AIAA Journal \textbf{45}(1), 257--270 (2007).
\newblock \doi{10.2514/1.24662}

\bibitem{Nuclear_Detonations}
K.~Terao, \emph{Irreversible Phenomena: Ignitions, Combustion and Detonation Waves} (Springer Berlin Heidelberg, Berlin, Heidelberg, 2007), chap. Industrial Applications of Detonation Waves, pp. 307--387.
\newblock \doi{10.1007/978-3-540-49901-5_12}

\bibitem{hondaSheetMetalForming1999}
A.~Honda, M.~Suzuki, Sheet metal forming by using gas imploding detonation.
\newblock Journal of Materials Processing Technology \textbf{85}(1), 198--203 (1999).
\newblock \doi{10.1016/S0924-0136(98)00317-3}

\bibitem{Perry_Kantrowitz}
R.W. Perry, A.~Kantrowitz, The production and stability of converging shock waves.
\newblock Journal of Applied Physics \textbf{22}(7), 878--886 (1951).
\newblock \doi{10.1063/1.1700067}

\bibitem{Saito_Glass}
T.~Saito, I.I. Glass, Temperature measurements at an implosion focus.
\newblock Proceedings of the Royal Society of London. A. Mathematical and Physical Sciences \textbf{384}(1786), 217--231 (1982).
\newblock \doi{10.1098/rspa.1982.0156}

\bibitem{Kjellander}
M.~Kjellander, N.~Tillmark, N.~Apazidis, Thermal radiation from a converging shock implosion.
\newblock Physics of Fluids \textbf{22}(4), 046102 (2010).
\newblock \doi{10.1063/1.3392769}

\bibitem{Shock_focusing_book}
N.~Apazidis, V.~Eliasson, \emph{Shock Focusing Phenomena, High Energy Density Phenomena and Dynamics of Converging Shocks} (Springer Cham, 2018).
\newblock \doi{10.1007/978-3-319-75866-4}

\bibitem{leeCylindricalImplodingShock1965}
J.H. Lee, B.H.K. Lee, Cylindrical imploding shock waves.
\newblock The Physics of Fluids \textbf{8}(12), 2148--2152 (1965).
\newblock \doi{10.1063/1.1761173}

\bibitem{Whitham1999}
G.B. Whitham, \emph{Linear and Nonlinear Waves} (John Wiley \& Sons, Ltd, 1999), chap. Shock Dynamics, pp. 263--338.
\newblock \doi{10.1002/9781118032954.ch8}

\bibitem{knystautasSparkInitiationConverging1967}
R.~Knystautas, J.H. Lee, Spark initiation of converging detonation waves.
\newblock AIAA Journal \textbf{5}(6), 1209--1211 (1967).
\newblock \doi{10.2514/3.4171}

\bibitem{knystautasDiagnosticExperimentsConverging1969}
R.~Knystautas, B.H.K. Lee, J.H.S. Lee, Diagnostic experiments on converging detonations.
\newblock The Physics of Fluids \textbf{12}(5), 165--168 (1969).
\newblock \doi{10.1063/1.1692602}

\bibitem{Lee_3}
R.~Knystautas, J.H. Lee, Experiments on the stability of converging cylindrical detonations.
\newblock Combustion and Flame \textbf{16}(1), 61--73 (1971).
\newblock \doi{10.1016/S0010-2180(71)80012-3}

\bibitem{ahlborn}
B.~Ahlborn, J.P. Huni, Stability and space-time measurements of concentric detonations.
\newblock AIAA Journal \textbf{7}(6), 1191--1192 (1969).
\newblock \doi{10.2514/3.5310}

\bibitem{Terao_1}
K.~Terao, Experimental study on cylindrical and spherical implosions.
\newblock Japanese Journal of Applied Physics \textbf{22}, 446 (1983).
\newblock \doi{10.1143/JJAP.22.446}

\bibitem{Terao_2}
K.~Terao, H.G. Wagner, Experimental study on spherically imploding detonation waves.
\newblock Shock Waves \textbf{1}(1), 27--34 (1991).
\newblock \doi{10.1007/BF01414865}

\bibitem{Terao_3}
K.~Terao, H.~Akaba, H.~Shiraishi, Spherically imploding detonation waves initiated by two-step divergent detonation.
\newblock Shock Waves \textbf{4}, 187--193 (1995).
\newblock \doi{10.1007/BF01414984}

\bibitem{sultanovFormationConvergingDetonation2023}
V.G. Sultanov, S.V. Dudin, V.A. Sosikov, S.I. Torunov, E.V. Vasilyunok, A.V. Razmyslov, D.Y. Rapota, Formation of a converging detonation wave with reverse front curvature.
\newblock Combustion, Explosion, and Shock Waves \textbf{59}(4), 516--525 (2023).
\newblock \doi{10.1134/S0010508223040159}

\bibitem{dudinExperimentalInvestigationCylindrical2016a}
S.V. Dudin, V.A. Sosikov, S.I. Torunov, Experimental investigation of cylindrical detonation wave.
\newblock Journal of Physics: Conference Series \textbf{774}, 012074 (2016).
\newblock \doi{10.1088/1742-6596/774/1/012074}

\bibitem{sosikovSmoothingFrontDetonation2019a}
V.A. Sosikov, S.I. Torunov, S.V. Dudin, Smoothing the front of the detonation wave in experiments with multipoint initiation.
\newblock Journal of Physics: Conference Series \textbf{1147}(1), 012027 (2019).
\newblock \doi{10.1088/1742-6596/1147/1/012027}

\bibitem{radulescu2018}
M.I. Radulescu, B.~Borzou, Dynamics of detonations with a constant mean flow divergence.
\newblock Journal of Fluid Mechanics \textbf{845}, 346--377 (2018).
\newblock \doi{10.1017/jfm.2018.244}

\bibitem{Xiao2020a}
Q.~Xiao, M.I. Radulescu, Dynamics of hydrogen–oxygen–argon cellular detonations with a constant mean lateral strain rate.
\newblock Combustion and Flame \textbf{215}, 437--457 (2020).
\newblock \doi{10.1016/j.combustflame.2020.01.041}

\bibitem{Xiao2020b}
Q.~Xiao, M.I. Radulescu, Role of instability on the limits of laterally strained detonation waves.
\newblock Combustion and Flame \textbf{220}, 410--428 (2020).
\newblock \doi{10.1016/j.combustflame.2020.06.040}

\bibitem{Xiao2021}
Q.~Xiao, A.~Sow, B.M. Maxwell, M.I. Radulescu, Effect of boundary layer losses on 2d detonation cellular structures.
\newblock Proceedings of the Combustion Institute \textbf{38}(3), 3641--3649 (2021).
\newblock \doi{10.1016/j.proci.2020.07.068}

\bibitem{yang2021}
H.~Yang, M.I. Radulescu, Dynamics of cellular flame deformation after a head-on interaction with a shock wave: Reactive richtmyer–meshkov instability.
\newblock Journal of Fluid Mechanics \textbf{923}, A36 (2021).
\newblock \doi{10.1017/jfm.2021.594}

\bibitem{MATLAB}
{The MathWorks Inc.}, \emph{Image Processing Toolbox, Version: 11.6 (R2022b)} (2022).
\newblock \urlprefix\url{https://www.mathworks.com/help/images/}

\bibitem{Manzhalei}
V.I. Manzhalei, V.V. Mitrofanov, V.A. Subbotin, Measurement of inhomogeneities of a detonation front in gas mixtures at elevated pressures.
\newblock Combustion, Explosion and Shock Waves \textbf{10}(1), 89--95 (1974).
\newblock \doi{10.1007/BF01463793}

\bibitem{desbordesCriticalDiameterDiffraction1985}
D.~Desbordes, M.~Vachon, \emph{Critical Diameter of Diffraction for Strong Plane Detonations}, in \emph{Proceedings of the 10th International Colloquium on the Dynamics of Explosions and Reactive Systems}, \emph{Progress in Astronautics and Aeronautics}, vol. 106 (1985), pp. 131--143.
\newblock \doi{10.2514/5.9781600865800.0131.0143}

\bibitem{desbordesTransmissionOverdrivenPlane1987}
D.~Desbordes, \emph{Transmission of Overdriven Plane Detonations: Critical Diameter as a Function of Cell Regularity and Size}, in \emph{Proceedings of the 11th International Colloquium on the Dynamics of Explosions and Reactive Systems}, \emph{Progress in Astronautics and Aeronautics}, vol. 114 (1987), pp. 170--185.
\newblock \doi{10.2514/5.9781600865886.0170.0185}

\bibitem{CEA}
S.~Gordon, B.J. McBride, Computer program for calculation of complex chemical equilibrium compositions and applications.
\newblock Tech. Rep. NASA-TR-1311, NASA Lewis Research Center (1996).
\newblock \urlprefix\url{https://ntrs.nasa.gov/citations/19950013764}

\bibitem{fayTwoDimensionalGaseousDetonations1959}
J.A. Fay, Two-dimensional gaseous detonations: Velocity deficit.
\newblock The Physics of Fluids \textbf{2}(3), 283--289 (1959).
\newblock \doi{10.1063/1.1705924}

\bibitem{Neemeh}
R.A. Neemeh, \emph{The Propagation and Stability of Converging Cylindrical Shocks in Narrow Cylindrical Chambers}, in \emph{Proceedings of the 18th International Symposium on Shock Waves}, ed. by K.~Takayama (1991), pp. 273--278.
\newblock \doi{10.1007/978-3-642-77648-9_38}

\bibitem{Mirels}
H.~Mirels, Attenuation in a shock tube due to unsteady-boundary-layer action.
\newblock Tech. Rep. NACA-TR-1333, NACA (1957).
\newblock \urlprefix\url{https://ntrs.nasa.gov/citations/19930092322}

\bibitem{Ng}
H.D. Ng, J.~Wang, J.H.S. Lee, \emph{Velocity Deficits in Thin Channels for a Cylindrically Expanding Detonation}, in \emph{Proceedings of the 25th International Colloquium on the Dynamics of Explosions and Reactive Systems} (2015), ICDERS, p. paper no. 223.
\newblock \urlprefix\url{http://www.icders.org/ICDERS2015/abstracts/ICDERS2015-223.pdf}

\end{thebibliography}

\end{document}